# HMS-VesselNet: Hierarchical Multi-Scale Attention Network with Topology-Preserving Loss for Retinal Vessel Segmentation

Amarnath R

*United Institute of Technology, Coimbatore, India*
**Date:** March 20, 2026
**Correspondence:** amarnathresearch@gmail.com

## Abstract

Retinal vessel segmentation methods based on standard overlap losses tend to miss thin peripheral vessels because these structures occupy very few pixels and have low contrast against the background. To address this, we propose HMS-VesselNet, a hierarchical multi-scale network that processes fundus images across four parallel branches at different resolutions and combines their outputs using learned fusion weights. The training loss combines Dice, binary cross-entropy, and centerline Dice to jointly optimize area overlap and vessel continuity. Hard example mining is applied from epoch 20 onward to concentrate gradient updates on the most difficult training images. Tested on 68 images from DRIVE, STARE, and CHASE_DB1 using 5-fold cross-validation, the model achieves a mean Dice of 88.72±0.67%, Sensitivity of 90.78±1.42%, and AUC of 98.25±0.21%. In leave-one-dataset-out experiments, AUC remains above 95% on each unseen dataset. The largest improvement is in the recall of thin peripheral vessels, which are the structures most frequently missed by standard methods and most critical for early detection of diabetic retinopathy.

## Highlights

- HMS-VesselNet architecture uses four-scale hierarchical Attention U-Nets with learned fusion weights favoring the 256x256 branch.
- Centerline Dice loss combined with deep supervision directly penalizes breaks in vessel continuity to improve topological accuracy.
- Hard example mining from epoch 20 oversamples difficult thin-vessel images at a 3× ratio to sustain the gradient signal.
- Achieved 90.78% mean Sensitivity and 88.72% Dice score across DRIVE, STARE, and CHASE_DB1 benchmarks.
- Leave-one-dataset-out (LODO) evaluation maintains AUC above 95%, confirming robust cross-dataset generalization.

## 1. Introduction

Diabetic retinopathy affects approximately 103 million people globally and is the leading cause of preventable vision loss [1]. Early detection depends on analyzing fundus images, where changes in vessel caliber, tortuosity, and branching patterns can precede symptomatic vision loss by several years [2, 3]. In large screening programs, tracking these changes requires automated segmentation of retinal vessels. Any missed or incorrect segmentation can directly affect clinical decisions, making segmentation accuracy a critical requirement.

However, segmentation accuracy is not equal for all vessels in the fundus image. Large vessels near the optic disc have well-defined boundaries and are relatively easier to segment, with overall Dice scores above 90% reported in recent evaluations [4, 5, 6]. Thin peripheral vessels are much harder to detect. They are often less than 2 pixels wide, have low contrast against the background, and appear in complex branching regions, as illustrated in Fig. 1. Sensitivity for these vessels frequently drops below 80% [7]. This is clinically important because changes in the caliber, tortuosity, and branching structure of small peripheral arterioles and venules are early indicators of microvascular damage in diabetic retinopathy, detectable years before symptomatic vision loss [2, 3]. Improving the segmentation of these fine structures is therefore a clinically meaningful objective.

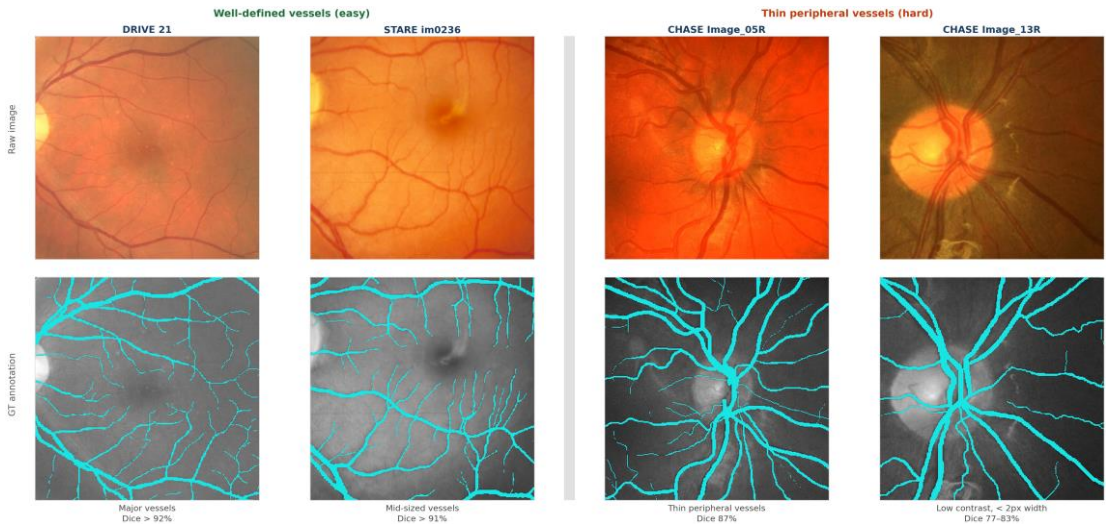

Fig. 1. Representative fundus image patches illustrating the range of segmentation difficulty. Left two patches show major and mid-sized vessels from DRIVE and STARE, where vessel boundaries are well defined and contrast is high. Right two patches show thin peripheral vessels from CHASE_DB1, where vessel width falls below 2 pixels and contrast is significantly lower.

Detecting these fine vessel structures is difficult for two main reasons related to standard loss functions. First, pixel-level losses like Dice and cross-entropy treat every pixel with equal weight. As a result, a thin vessel of 200 pixels creates a much weaker gradient than a major vessel of 2000 pixels, allowing larger structures to dominate the training. Second, these losses are topologically insensitive. Since they measure area overlap, a single break in a vessel centerline has minimal impact on the total loss value. Therefore, in morphological analysis, a broken vessel connection is a segmentation failure, even if the surrounding region is predicted correctly.

Several approaches have been proposed to address these limitations. Multi-scale architectures process inputs at different resolutions to capture fine vessel features [7], and attention gates suppress irrelevant background activations in the decoder [8, 12]. While these methods improve detection of small structures, they do not directly enforce vessel continuity. clDice loss addresses this by penalizing breaks in the vessel skeleton, but whether combining it with multi-scale fusion and hard example mining can improve segmentation performance beyond what each achieves individually has not been studied.

In this paper, we introduce HMS-VesselNet, a multi-scale segmentation network designed to address these limitations. The model uses four Attention U-Net branches operating at different input resolutions in parallel, with learned weights to merge their outputs. The training loss combines Dice, cross-entropy, and clDice to optimize area overlap and vessel continuity. Hard example mining is incorporated into the training loop to give more attention to difficult images. Evaluated on 68 images across DRIVE, STARE, and CHASE_DB1 using 5-fold cross-validation, the model achieves a mean Dice of 88.72%, Sensitivity of 90.78%, and AUC of 98.25%.

The contributions of this work are as follows.

1. A hierarchical fusion architecture using four scales with learned weights. We find the 256×256 branch consistently carries the most weight, indicating that it is the most reliable scale for capturing vessel structure.
2. A composite loss function that targets both area overlap and centerline continuity through deep supervision at every branch.
3. The integration of hard example mining within the training loop, supported by ablation studies that isolate its contribution to Sensitivity from that of the architecture and loss function.
4. A leave-one-dataset-out (LODO) evaluation across three benchmark datasets to assess how well the model generalizes to unseen acquisition conditions.

The rest of the paper is organized as follows. Section 2 reviews related work in retinal segmentation and multi-scale learning. Section 3 details the architecture and loss functions. Section 4 covers the datasets and implementation. Results and ablation studies are reported in Section 5, followed by a discussion of limitations and future directions in Section 6.

## 2. Related Work

Retinal vessel segmentation has been approached in several directions over the past two decades. This section reviews the four areas most directly relevant to HMS-VesselNet namely general segmentation methods including classical and deep learning approaches, multi-scale architectures for handling the range of vessel widths within a single image, attention mechanisms for suppressing irrelevant background features, and topology-preserving losses for maintaining vessel continuity. Each subsection closes by identifying the limitation that the proposed HMS-VesselNet is designed to address.

### 2.1. Retinal Vessel Segmentation

Early approaches to retinal vessel segmentation were built around explicit signal processing operations such as matched filters [10], vessel tracking [38], and morphological processing [3]. These methods perform reasonably on high-contrast major vessels but require careful parameter tuning and degrade on thin peripheral vessels and pathological images where contrast is inconsistent. Convolutional neural networks (CNN) improved this substantially, replacing fixed signal processing pipelines with learned representations that generalize better across imaging conditions. Early CNN-based methods using encoder-decoder architectures on DRIVE and STARE reported improvements over classical approaches in both Dice and sensitivity [7, 12]. The introduction of fully convolutional networks and encoder-decoder architectures allowed the model to learn spatial context rather than per-pixel features in isolation [14]. U-Net [15] and its variants became the standard backbone for this task, with residual connections [16] and dense connectivity [17] further improving feature reuse across encoder and decoder layers.

More recent work has focused on improving thin vessel recall specifically. Wang et al. [18] introduced a dual-branch network separating thick and thin vessel pathways. Jin et al. [19]

introduced deformable convolutions that adapt receptive fields to vessel shape, improving detection of curved and thin vessels. Transformer-based architectures have also been applied to retinal vessel segmentation, with self-attention mechanisms providing global context that convolutional encoders lack [5, 6, 13]. Hybrid designs combining convolutional encoders with transformer decoders have shown particular promise, though they require substantially more training data and compute than pure convolutional approaches. Despite these advances, sensitivity on thin peripheral vessels remains below 80% in most reported evaluations [7], and cross-dataset generalization is infrequently tested. Most papers report results on a single benchmark, making it difficult to assess whether gains reflect genuine improvements or adaptation to dataset-specific characteristics. HMS-VesselNet addresses this by evaluating all three benchmarks using LODO testing.

## 2.2. Multi-Scale Architecture

Retinal vessels span a wide range of spatial scales within a single image. Vessels near the optic disc may be several pixels wide, while terminal branches near the periphery can be a single pixel wide. A network operating at a fixed resolution cannot simultaneously optimize for both, because the receptive field and feature resolution suited major vessels differ fundamentally from those required for thin ones.

Feature pyramid networks [20] address this by aggregating features from multiple decoder levels within a single forward pass. Applied to vessel segmentation, this improves thin vessel recall relative to single-scale baselines [21, 12]. However, since these approaches process the image at a single input resolution, their fine-resolution feature maps are derived from a downsampled representation, which limits spatial precision for thin structures.

Parallel multi-scale architectures [7] avoid this by processing the input independently at each resolution. MS-NFN [7] and similar designs pass the same image through separate branches at different scales, preserving full spatial detail rather than recovering it from a bottleneck. The remaining limitation is that most such designs use fixed fusion weights, either simple averaging or manually chosen schemes. Because the optimal contribution of each scale varies with dataset characteristics and acquisition conditions, HMS-VesselNet learns the fusion weights jointly with the segmentation objective, allowing the model to discover which scale is most informative for this specific task.

## 2.3. Attention Mechanisms in Segmentation

Skip connections in encoder-decoder networks preserve high-resolution spatial detail by passing encoder feature maps directly to the decoder. In practice, however, they also pass irrelevant background activations that compete with vessel features in the decoder, suppressing detection of thin structures whose signal is already weak.

Attention gates [8] address this by computing a spatial coefficient for each skip connection based on the current decoder state, suppressing irrelevant regions before they reach the decoder. On retinal vessel benchmarks, Attention U-Net consistently improves sensitivity on thin vessels compared to standard U-Net at equivalent depth and parameter count [23, 24]. Channel attention mechanisms such as squeeze-and-excitation blocks [22] offer complementary gains by reweighting feature channels globally, and several papers report additive improvements when both are combined [23, 24]. However, the individual contribution of each component is rarely isolated in ablation studies. HMS-VesselNet uses spatial attention gates at every skip connection without adding channel attention, keeping the ablation analysis tractable and the contribution of each design choice interpretable.

## 2.4. Topology-Preserving Losses

Standard overlap losses such as Dice are insensitive to topological errors in predicted segmentations. A prediction that disconnects a vessel at a single point incurs only a small area of penalty, yet the consequence is a broken vessel graph that cannot be used for morphological analysis. This gap between the optimization objective and clinical utility has motivated work on topology-aware training objectives.

Clough et al. [25] proposed a loss based on persistent homology that directly penalizes topological differences between predicted and ground-truth segmentations. Hu et al. [26] introduced a more efficient alternative using critical points in the distance transform. Both are theoretically well-motivated but add considerable computational overhead and have seen limited adoption in retinal vessel segmentation pipelines.

Shit et al. [9] introduced clDice, which computes Dice loss on the soft skeleton of the predicted and ground-truth masks rather than their full area. Soft skeletonization is implemented through differentiable morphological operations, making clDice fully compatible with standard backpropagation at very low additional computational cost. By measuring agreement along the centerline rather than across the vessel area, clDice directly penalizes breaks in vessel continuity. On tubular structure segmentation tasks including retinal vessels, clDice improves connectivity metrics and sensitivity compared to Dice alone [9]. HMS-VesselNet integrates clDice to ensure that sensitivity gains do not come at the cost of vessel fragmentation.

## 3. Methodology

Fig. 2 provides an overview of HMS-VesselNet. The pipeline consists of four stages namely input preprocessing, four parallel Attention U-Net branches operating at different resolutions, learned scale fusion, and a composite loss with deep supervision. Each stage is detailed below.

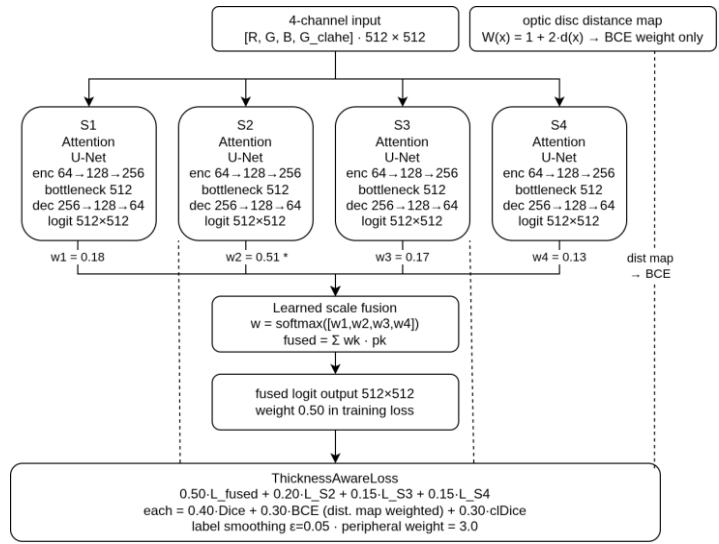

*Fig. 2. Overall architecture of HMS-VesselNet.*

## 3.1. Input Preprocessing

Raw fundus images are provided as 8-bit RGB. The green channel carries the highest vessel-to-background contrast because hemoglobin absorption peaks in the green wavelength range, but its absolute brightness and contrast vary across imaging devices and acquisition conditions. To address this, we construct a 4-channel input tensor through two parallel paths.

The first path converts the image to LAB colour space, enhances the L-channel using CLAHE [27] (clip limit 2.0, tile grid 8×8), and converts the result back to RGB, producing channels 1 to 3. The second path independently enhances the green channel of the original unmodified RGB image using CLAHE (clip limit 3.0, tile grid 8×8), producing channel 4. Concatenating both paths into a 4×H×W tensor provides the network with a contrast-normalized color signal alongside a dedicated vessel contrast channel.

An optic disc distance map is computed for each image to serve as a pixel-level weighting term in the loss function. The disc is localised by applying Gaussian blur (51×51 kernel) to the LAB L-channel followed by peak detection [34]. A Euclidean distance transform then encodes each pixel's normalized distance from the disc center, denoted $d(x)$.

This map weights the BCE loss term to assign progressively higher gradients toward peripheral pixels, where thin vessels are concentrated. The weight $W(x)$ increases linearly from 1.0 at the disc centre to 3.0 at the periphery:

$$W(x) = 1.0 + 2.0 \times d(x) \tag{1}$$

A linear function was chosen for interpretability. The upper bound of 3.0 was determined empirically across {2.0, 3.0, 4.0, 5.0}. Values above 3.0 caused over-segmentation in low-contrast peripheral regions without improving thin vessel recall. This distance map is used exclusively during loss computation and does not enter the network as an input channel.

## 3.2. Network Architecture

HMS-VesselNet runs four Attention U-Net branches [8] in parallel, with a single branch illustrated in Fig. 3. Each branch receives the 4-channel input resized to its native resolution, namely, S1 at 512×512, S2 at 256×256, S3 at 128×128, and S4 at 64×64, covering coarse-to-fine spatial scales across the vascular hierarchy. S1 preserves fine vessel detail, S2 captures branching patterns, S3 encodes vessel topology context, and S4 captures global vascular layout.

A four-scale design was preferred over three scales based on preliminary experiments showing consistent Sensitivity improvement with the S4 branch. A fifth scale was excluded because at 32×32, thin vessel structures are no longer reliably preserved after downsampling.

All branches share an identical encoder comprising three blocks with channel dimensions 64, 128, and 256. Each block contains two 3×3 convolutions with batch normalisation and ReLU activation, followed by MaxPool2d downsampling. A bottleneck block at 512 channels with dropout rate 0.4 follows the third encoder block. At each decoder level, an attention gate uses the upsampled decoder feature as a gating signal to reweight the corresponding encoder skip connection before concatenation, suppressing background activations. Weights are initialised using the Kaiming normal scheme [28].

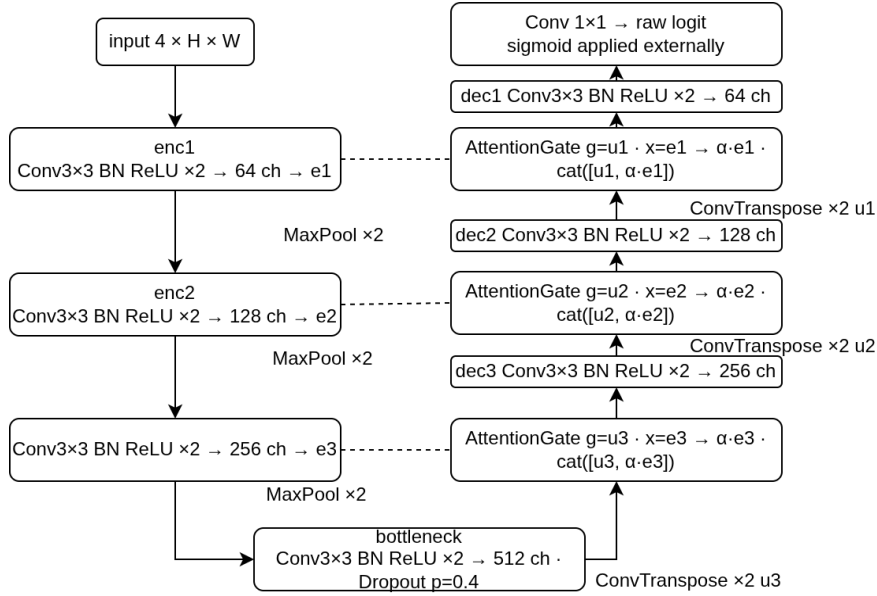

*Fig. 3. Architecture of a single Attention U-Net branch.*

The output logit maps from all four branches are bilinearly upsampled to 512×512 and combined through weighted sum. The fusion weights w = [w1, w2, w3, w4] are trainable scalar parameters passed through softmax to ensure they remain non-negative and sum to one.

$$P_{fused} = \sum_{k=1}^{4} soft\max(w) \cdot upsample(L_k) \quad (2)$$

where $L_k$ is the output logit map of branch k and upsample denotes bilinear interpolation to 512×512.

Initial weights are set to [0.40, 0.25, 0.20, 0.15], reflecting a prior toward higher-resolution branches. Equal initialisation at [0.25, 0.25, 0.25, 0.25] was tested but produced slower convergence, requiring the network to first discount coarser branches before S1 and S2 weights reached useful ranges. The descending initialisation provides a warm start while still permitting substantial adaptation, as confirmed by the converged S2 dominance reported in Section 5.4. During inference, sigmoid activation is applied to the fused logit to obtain the final probability map.

## 3.3. Loss Function

Standard overlap losses alone are insufficient for retinal vessel segmentation because they are insensitive to class imbalance, topological continuity, and the spatial distribution of thin vessels across the image. To address these three limitations, the training objective L is defined as a weighted combination of three complementary terms such as Dice loss ($L_{Dice}$) for area overlap, binary cross-entropy ($L_{BCE}$) with peripheral weighting for spatial gradient control, and centerline Dice ($L_{clDice}$) for topological continuity.

$$L = 0.40 \times L_{Dice} + 0.30 \times L_{BCE} + 0.30 \times L_{clDice} \quad (3)$$

The weights were selected by grid search over {0.33, 0.40, 0.50} for $L_{Dice}$, with the remainder split equally between $L_{BCE}$ and $L_{clDice}$. The chosen values maximised mean Sensitivity on the Fold 1 validation set and were fixed for all subsequent experiments.

$L_{Dice}$ measures volumetric overlap between the predicted probability map and the binary ground truth mask. It is insensitive to class imbalance and provides stable gradients when foreground pixels are sparse, which is the typical condition in fundus vessel segmentation where vessels occupy 10% to 15% of image pixels.

$$L_{Dice} = 1 - \frac{2 \sum p \cdot g + \varepsilon}{\sum p + \sum g + \varepsilon} \quad (4)$$

where 'p' denotes predicted probabilities, 'g' denotes binary ground truth labels, and 'ε' is a smoothing constant preventing division by zero.

$L_{BCE}$ uses label smoothing with ε = 0.05 to prevent overconfidence on easy background pixels and improve calibration in dense prediction tasks [29]. The optic disc distance map is applied as a pixel-level weighting map, with weights varying continuously from 1.0 at the disc center to 3.0 at the periphery, progressively increasing gradient contribution toward peripheral pixels where thin vessels are concentrated. The Dice and clDice terms are unaffected.

$$L_{BCE} = -mean[W(x) \cdot [gs \log(p) + (1-gs) \log(1-p)]] \quad (5)$$

where gs = (1−0.05) × g + 0.025 are the smoothed labels and W(x) = 1.0 + 2.0 × d(x) is the peripheral weighting map, with d(x) the normalised Euclidean distance from the optic disc centre in [0,1].

$L_{clDice}$ [9] computes Dice loss on the soft skeletons of the predicted and ground truth masks rather than their full area. Soft skeletonization is implemented through five iterations of differentiable morphological erosion and opening, making it fully compatible with standard backpropagation. By penalizing disagreements along the vessel centerline, $L_{clDice}$ directly discourages predictions that break vessel continuity. This represents a topological failure that area-based losses largely ignore.

$$L_{clDice} = 1 - \frac{2 \cdot Tprec \cdot Tsens}{Tprec \cdot Tsens + \varepsilon} \quad (6)$$

where,

$$T_{prec} = \frac{\sum skel(p) \cdot g + \varepsilon}{\sum skel(p) + \varepsilon} \qquad T_{sens} = \frac{\sum skel(g) \cdot p + \varepsilon}{\sum skel(g) + \varepsilon}$$

and skel(·) denotes soft skeletonization over five morphological iterations.

Deep supervision is applied to branches S2, S3, and S4 directly. Branch S1 is not separately supervised as it contributes to the fused output, which already receives the highest supervision weight and serves as the primary optimisation target. Each auxiliary branch is supervised at its native resolution using ground truth downsampled to match, with the same composite loss. The total supervised loss is:

$$L_{total} = 0.50 \times L_{fused} + 0.20 \times L_{S2} + 0.15 \times L_{S3} + 0.15 \times L_{S4} \quad (7)$$

The fused output receives the highest weight as it benefits from all four branches. Auxiliary weights decrease with resolution, reflecting the reduced spatial detail available at coarser scales.

### 3.4. Training Procedure

The network is trained using AdamW [30] with an initial learning rate of $1\times10^{-3}$ and weight decay of $1\times10^{-4}$. The learning rate follows a cosine annealing [31] schedule with warm restarts, with an initial cycle length of 40 epochs, a cycle multiplier of 2, and a minimum learning rate of $1\times10^{-6}$. Training uses a batch size of 2 with early stopping at a patience of 30 epochs, monitoring validation Dice on the held-out fold. Gradient clipping is applied at a maximum norm of 1.0 to stabilise training during cosine annealing restarts.

Data augmentation is applied at native resolution before resizing to 512×512. The spatial augmentation pipeline includes horizontal and vertical flips, 90-degree rotations, shift-scale-rotate (shift 0.1, scale 0.1, rotation up to 30 degrees), and elastic transformations (α=120, σ=6), following standard recommendations for medical image segmentation in the Albumentations library [37].

Photometric augmentation includes random brightness and contrast variation (±0.3), hue-saturation-value jitter, CLAHE [27] with clip limit 4.0, random gamma in the range 80 to 120, Gaussian noise, and Gaussian blur. The ±0.3 brightness and contrast range was chosen to cover the illumination differences observed across datasets, particularly between the Canon CR5 acquisitions in DRIVE and the hand-held Nikon NM-200D images in CHASE_DB1. The augmentation CLAHE clip limit of 4.0 is intentionally higher than the preprocessing limits of 2.0 and 3.0 to simulate aggressive contrast variation during training. Mixup [32] (α=0.2) is applied with probability 0.5, following standard practice for dense prediction tasks where higher alpha values blur vessel boundaries. This pipeline covers the imaging variability seen across the three datasets without introducing unrealistic artefacts.

Fig. 4 illustrates the hard example mining schedule, which activates from epoch 20 onward. At each epoch, every training image is scored by 1 Dice on the unaugmented image, ensuring stable and reproducible difficulty scores across epochs. The top 15% of images by this score are oversampled at 3× using a weighted random sampler. The 15% threshold concentrates oversampling on genuinely difficult images while retaining sufficient batch diversity. Values below 10% produced overly homogeneous batches, while values above 20% diluted the effect by including moderately difficult images alongside the hardest ones. The 3× rate balances amplifying the gradient signal on hard images against overfitting to a small fixed subset. Activation is delayed until epoch 20 to allow the network to reach a stable state where per-image Dice scores reflect intrinsic difficulty rather than initialisation effects. In practice, the consistently hardest images are peripheral thin-vessel cases from CHASE_DB1.

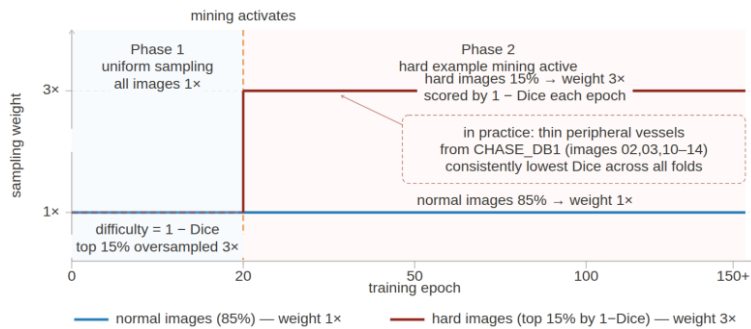

*Fig. 4. Hard example mining schedule.*

At test time, 8-orientation test-time augmentation is applied by averaging predictions across four flip combinations (original, horizontal flip, vertical flip, and both) and two rotation variants (90 and 270 degrees). This scheme was preferred over 4-orientation TTA because the additional rotation variants consistently reduced variance in Sensitivity on CHASE_DB1 without a meaningful increase in inference time. The averaged probability map is thresholded at 0.5 to produce the final binary segmentation. TTA is applied exclusively during final testing and omitted during validation for early stopping, to keep validation computationally tractable.

## 4. Experiments

### 4.1. Datasets

Three publicly available retinal fundus datasets are used in this study, which are summarised in Table 1. They differ in camera type, image resolution, subject population, and pathology range, providing a more demanding evaluation than any single dataset alone and requiring the model to generalize across acquisition conditions rather than overfit to a single imaging setup. All images are 8-bit RGB acquired in a 45-degree field of view. Vessel pixels constitute approximately 10% to 15% of total image pixels across all three datasets, reflecting the class imbalance that motivated the loss function design in Section 3.3. Images from each dataset are resized to 512×512 using bilinear interpolation before being passed to the network.

*Table 1:* Summary of the three datasets used in this study. All images are 8-bit RGB with manual vessel annotations provided by the respective dataset authors.

| Dataset | Images | Resolution | Camera | Population | Pathology |
|---|---|---|---|---|---|
| **DRIVE [33]** | 20 | 565×584 | Canon CR5 | Adults, 25-90 yrs | Diabetic retinopathy screening |
| **STARE [10]** | 20 | 700×605 | TopCon TRV-50 | Mixed | Normal and pathological |
| **CHASE_DB1 [35]** | 28 | 1280×960 | Nikon NM-200D | School children | Varied image quality |
| **Total** | **68** | — | — | — | — |

DRIVE [33] consists of 20 fundus images collected during a diabetic retinopathy screening programme in the Netherlands, acquired with a Canon CR5 non-mydriatic camera. Manual annotations are provided by two independent observers. The first observer annotation is used as ground truth, which is the standard choice in published comparisons on this dataset [4, 12].

STARE [10] contains 20 images acquired with a TopCon TRV-50 camera, covering a wider range of pathological conditions than DRIVE, including papilloedema, arteriovenous malformations, and choroidal neovascularization. Ten images are from normal subjects and ten from patients with retinal disease. Two manual annotations are available, however the first annotator is used as ground truth, consistent with the majority of published work on STARE [11, 12].

CHASE_DB1 [35] comprises 28 images from both eyes of 14 school children, acquired with a hand-held Nikon NM-200D camera. Hand-held acquisition introduces greater variability in illumination, focus, and field alignment compared to tabletop cameras. The combination of thin vessels and low image contrast makes this dataset consistently the hardest of the three for most methods. Two manual annotations are available where the first annotation is used as ground truth. Fig. 5 shows representative patches from all three datasets alongside their ground truth annotations, illustrating the differences in vessel contrast and image quality across datasets.

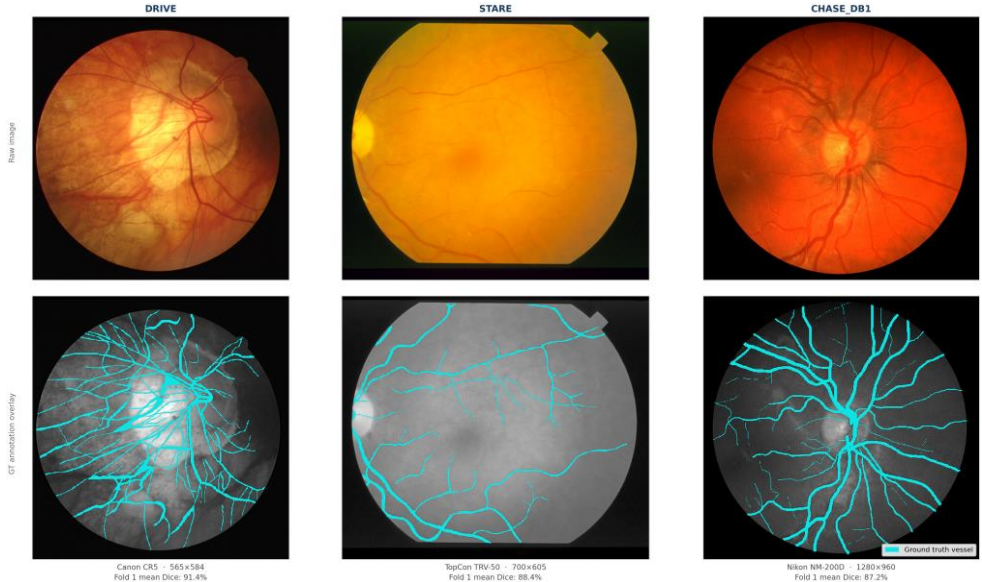

*Fig. 5. Representative images from DRIVE (left), STARE (centre), and CHASE_DB1 (right), with ground truth vessel annotations overlaid. The difference in vessel contrast, image uniformity, and thin vessel visibility is apparent across the three datasets. Mean Dice in Fold 1 is 91.4% on DRIVE, 88.4% on STARE, and 87.2% on CHASE_DB1.*

The HRF dataset [36] was excluded for a specific technical reason. HRF images have a native resolution of 3504×2336 pixels, requiring a downsampling ratio of approximately 6.9× to reach 512×512. This is substantially more aggressive than the ratios applied to the included datasets, namely 1.1× for DRIVE (565×584), 1.4× for STARE (700×605), and 2.5× for CHASE_DB1 (1280×960). At a 6.9× ratio, vessel widths of 2 to 3 pixels in the original HRF image would be reduced to sub-pixel representations at 512×512, effectively erasing the fine vessel structures that make HRF a challenging benchmark. Including HRF under these conditions would not constitute a fair evaluation and would render cross-dataset comparisons invalid. Future work with a pipeline designed natively for high-resolution inputs, or incorporating a super-resolution preprocessing step, would be the appropriate setting for HRF evaluation.

### 4.2. Evaluation Protocol

All results are reported under 5-fold stratified cross-validation with a fixed random seed (seed = 42) to ensure reproducibility of fold assignments. Folds are stratified by dataset so that each fold contains exactly 4 DRIVE images, 4 STARE images, and either 5 or 6 CHASE_DB1 images, giving held-out sets of 13 or 14 images per fold. Fig. 6 illustrates the fold structure. Training sets contain 54 or 55 images depending on the fold.

| Fold | Training set (54–55 images) | Held-out set (13–14 images) |
|---|---|---|
| F1 | DRIVE:16  STARE:16  CHASE:22 | DRIVE:4  STARE:4  CHASE:14 |
| F2 | DRIVE:16  STARE:16  CHASE:22 | DRIVE:4  STARE:4  CHASE:14 |
| F3 | DRIVE:16  STARE:16  CHASE:22 | DRIVE:4  STARE:4  CHASE:14 |
| F4 | DRIVE:16  STARE:16  CHASE:23 | DRIVE:4  STARE:4  CHASE:13 |
| F5 | DRIVE:16  STARE:16  CHASE:23 | DRIVE:4  STARE:4  CHASE:13 |

Total: 68 images [DRIVE: 20 · STARE: 20 · CHASE: 28]
Each image appears in held-out set exactly once

*Fig. 6. 5-fold stratified cross-validation scheme with exact image counts per fold.*

Within each fold, the held-out images serve a dual role. During training they are used as the validation set for early stopping, with Dice computed without test-time augmentation to keep validation computationally tractable. After training completes, the same held-out images are used as the test set with 8-orientation TTA applied. Because TTA consistently improves Dice by a small margin relative to no augmentation, the test Dice is slightly higher than the validation Dice used for early stopping. This is the source of the 0.07% mean validation-to-test gap reported in Section 5, not overfitting the validation set.

Performance is reported using four metrics. Let TP, FP, TN, and FN denote true positives, false positives, true negatives, and false negatives at the pixel level, where a positive prediction corresponds to a vessel pixel. All threshold-dependent metrics are computed on the binary segmentation map obtained by thresholding the averaged TTA probability map at 0.5. This threshold was retained as the default throughout all experiments rather than tuned per fold, to avoid introducing fold-specific optimism into the reported metrics.

The Dice coefficient measures overlap between the predicted and ground truth vessel masks.

$$Dice = \frac{2 \times TP}{2 \times TP + FP + FN} \tag{8}$$

Sensitivity measures the proportion of true vessel pixels correctly detected.

$$Sensitivity = \frac{TP}{(TP + FN)} \tag{9}$$

Specificity measures the proportion of true background pixels correctly rejected.

$$Specificity = \frac{TN}{(TN + FP)} \tag{10}$$

The area under the receiver operating characteristic curve (AUC) is computed from the predicted probability map before thresholding, by varying the decision threshold from 0 to 1 and recording sensitivity and 1 Specificity at each threshold. AUC measures ranking quality independently of the chosen threshold and is reported alongside threshold-dependent metrics to provide a complete picture of model performance.

Sensitivity is the primary reported metric. In retinal screening, a missed vessel pixel is more consequential than a false detection because false negatives correspond to undetected pathological change, whereas false positives can be identified by a grader during secondary review. Dice and AUC are reported as secondary metrics. Specificity is included to confirm that Sensitivity gains are not accompanied by a proportional increase in false positives.

Fig. 7 shows an example prediction from a CHASE_DB1 peripheral image alongside the ground truth and error map, with false negatives shown in red and false positives in blue. The distribution of errors is representative of the pattern observed across difficult cases: false negatives concentrated at thin peripheral branches, and false positives sparse and scattered. These thin peripheral branches are the primary target of the clDice loss and hard example mining described in Section 3.

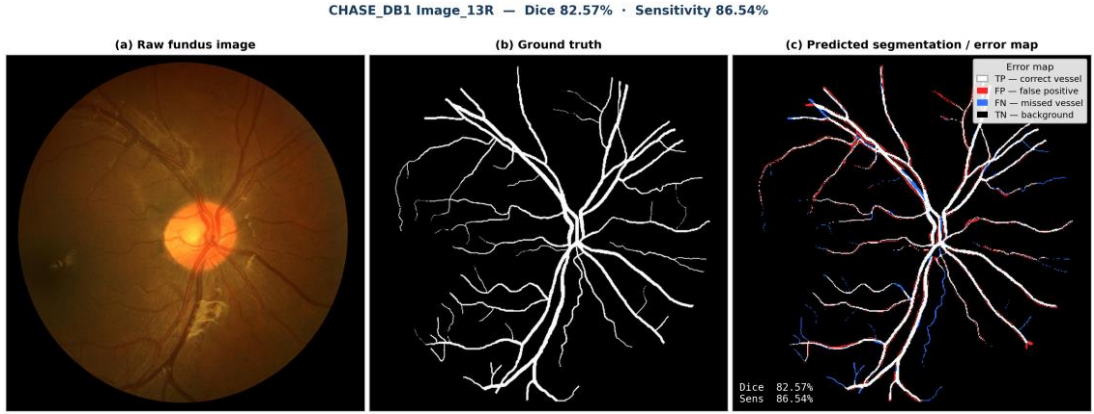

*Fig. 7. Example segmentation output on a CHASE_DB1 peripheral patch. (a) Raw fundus image. (b) Ground truth vessel annotation. (c) Predicted segmentation with error map: false negatives in red, false positives in blue.*

### 4.3. Leave-One-Dataset-Out Protocol

Cross-validation measures performance when training and test images share the same underlying distribution. It does not establish whether the model generalizes a completely unseen acquisition setup. To address this, we conducted three leave-one-dataset-out (LODO) experiments. In each experiment, one dataset is withheld entirely from training and used exclusively for evaluation, while the remaining two datasets provide all training images.

The three experiments are defined as follows: (a) LODO-DRIVE, training on STARE and CHASE_DB1, testing on all 20 DRIVE images (b) LODO-STARE, training on DRIVE and CHASE_DB1, testing on all 20 STARE images and (c) LODO-CHASE, training on DRIVE and STARE, testing on all 28 CHASE_DB1 images. Each experiment uses the exact same training procedure as the cross-validation runs, including hard example mining, data augmentation, and 8-orientation TTA at test time. No hyperparameter adjustment is made based on the held-out dataset in any of the three runs.

We compare the LODO results against the cross-validation mean as a baseline reference. A drop in the Dice score relative to cross-validation is expected because the training set is smaller and contains no examples from the test camera type. The primary objective of these experiments is to determine whether the AUC and Sensitivity metrics remain at a practically useful level when the model processes images from a completely unseen acquisition condition.

### 4.4. Implementation Details

The model is implemented in PyTorch 2.0. The total parameter count is approximately 124 million, comprising four Attention U-Net branches of approximately 31 million parameters each. Training uses mixed precision with automatic loss scaling to reduce memory usage. The batch size is restricted to 2 due to GPU memory constraints imposed by running four full-resolution branches simultaneously.

Training is conducted on an NVIDIA Tesla T4 GPU with 16 GB of memory within the Kaggle computational environment. Depending on the early stopping epoch, each cross-validation fold requires approximately 4 to 6 hours to reach convergence, resulting in a total compute time of approximately 25 hours across all five folds. Each LODO run requires a comparable duration.

Inference for a single image, including 8-orientation TTA, takes approximately 1.2 seconds on this hardware.

All three datasets are publicly available and were downloaded from their respective project pages without modification. Ground truth annotations are used exactly as provided by the dataset authors. To ensure full reproducibility, the source code, exact fold assignments, and trained model weights will be made publicly available on GitHub upon publication.

## 5. Results

### 5.1. Cross-Validation Results

Table 2 reports per-fold and mean results across all five folds under 8-orientation TTA. The model achieves a mean Dice of 88.72±0.67%, AUC of 98.25±0.21%, Sensitivity of 90.78±1.42%, and Specificity of 99.21±0.10%. The low standard deviation on AUC and Specificity indicates consistent ranking quality and false positive rate across folds. Specificity remains consistently high across folds because the background class dominates the image where vessel pixels constitute only 10% to 15% of total pixels, making correct rejection of background pixels straightforward regardless of fold composition. The clinically meaningful variation is therefore in Sensitivity. The wider standard deviation of 1.42 percentage points reflects the uneven distribution of hard CHASE_DB1 images across folds.

**Table 2.** *Per-fold and mean cross-validation results under 8-orientation TTA on the held-out test set. Val N is the number of held-out images per fold.*

| Fold | Val N | Dice (%) | AUC (%) | Sensitivity (%) | Specificity (%) | Best Epoch |
|---|---|---|---|---|---|---|
| 1 | 14 | 88.94 | 98.39 | 89.72 | 99.29 | 118 |
| 2 | 14 | 88.51 | 97.98 | 89.46 | 99.26 | 111 |
| 3 | 14 | 88.6 | 98.22 | 90.17 | 99.25 | 104 |
| 4 | 13 | 87.86 | 98.14 | 91.86 | 99.03 | 111 |
| 5 | 13 | 89.7 | 98.53 | 92.7 | 99.21 | 102 |
| **Mean±SD** | — | **88.72±0.67** | **98.25±0.21** | **90.78±1.42** | **99.21±0.10** | — |

Fig. 8 shows the training curves for all five folds, confirming stable convergence throughout. Best Epoch is the epoch at which early stopping triggered on validation Dice. All five folds converge stably between epochs 102 and 118, with cosine annealing warm restarts visible as periodic increases in learning rate. Fold 4 achieves the lowest Dice at 87.86%, corresponding to the highest concentration of hard peripheral CHASE_DB1 images in the held-out set (Image_10L, 11L, 13L, 14L, 14R). Fold 5 achieves the highest Dice at 89.70%, reflecting a held-out set weighted toward DRIVE images with clear vessel contrast.

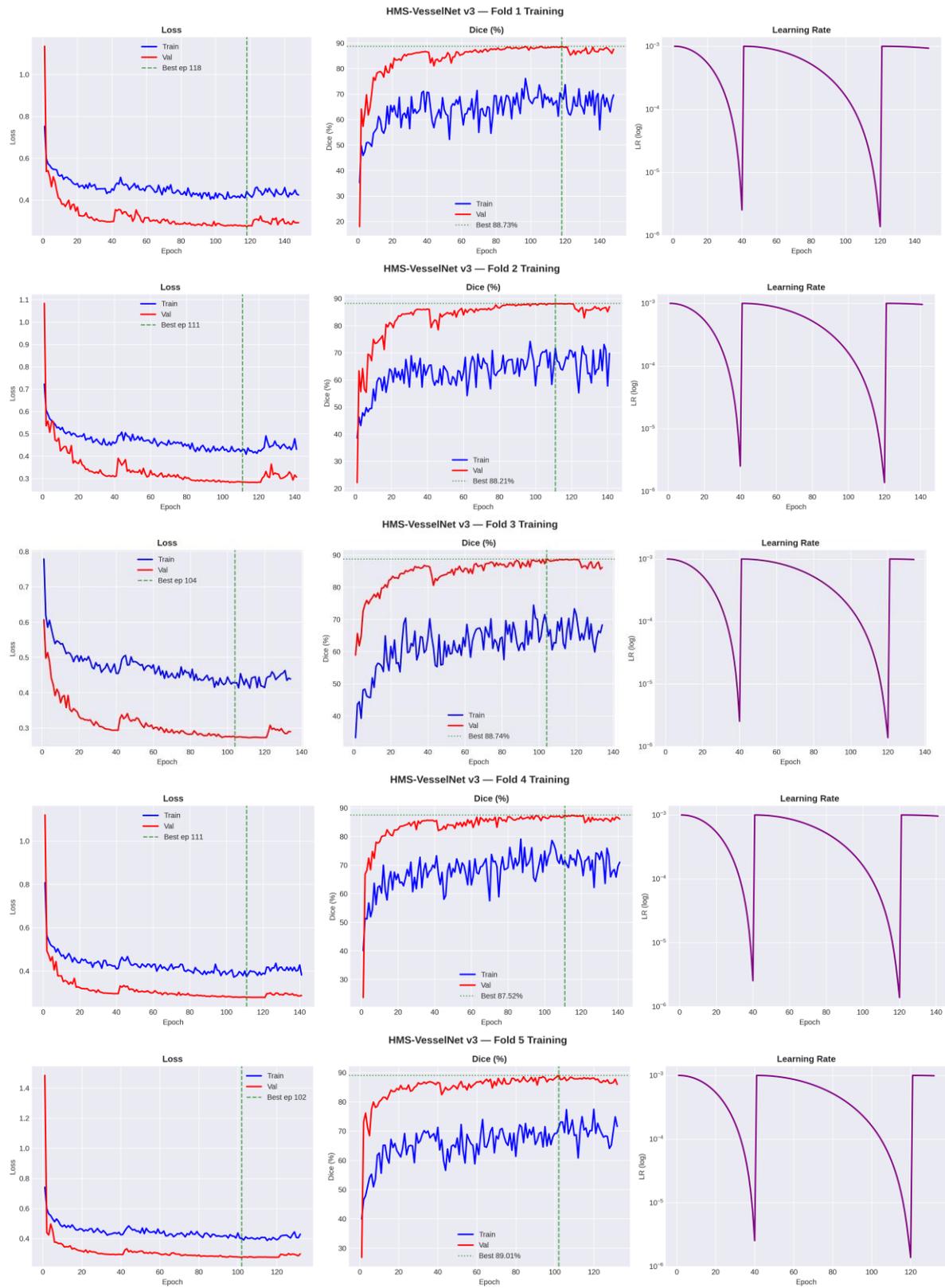

Fig. 8. Training curves for all five cross-validation folds showing training loss and validation Dice over epochs.

The mean validation-to-test gap across all five folds is 0.07%, confirming that the dual use of the held-out fold for both early stopping and final evaluation does not produce optimistic estimates. As described in Section 4.2, this gap reflects the marginal benefit of 8-TTA over no TTA rather than any overfitting to the validation set.

## 5.2. Per-Image Analysis

Table 3 reports per-image Dice and Sensitivity for all 14 images in the Fold 1 held-out set grouped by dataset. All metrics computed with 8-orientation TTA. Outlier rows are STARE im0240 (arteriovenous malformation) and CHASE_DB1 Image_13R (thin vessels, low contrast).

*Table 3. Per-image Dice and Sensitivity for all 14 held-out images in Fold 1 grouped by dataset.*

| Dataset | Image | Dice (%) | Sensitivity (%) |
|---|---|---|---|
| **DRIVE** | 21_training.tif | 92.04 | 95.66 |
| **DRIVE** | 22_training.tif | 90.92 | 88.83 |
| **DRIVE** | 36_training.tif | 92.11 | 87.98 |
| **DRIVE** | 38_training.tif | 90.55 | 93.12 |
| | **Mean** | **91.41** | **91.4** |
| **STARE** | im0001.ppm | 90.45 | 90.24 |
| **STARE** | im0002.ppm | 90.21 | 88.45 |
| **STARE** | im0240.ppm | 83.37 | 75.84 |
| **STARE** | im0291.ppm | 89.37 | 85.21 |
| | **Mean** | **88.35** | **84.94** |
| **CHASE** | Image_01L.jpg | 90.85 | 92.92 |
| **CHASE** | Image_05L.jpg | 89.15 | 91.28 |
| **CHASE** | Image_05R.jpg | 90.86 | 92.26 |
| **CHASE** | Image_07L.jpg | 88.34 | 91.18 |
| **CHASE** | Image_11R.jpg | 84.37 | 96.63 |
| **CHASE** | Image_13R.jpg | 82.57 | 86.54 |
| | **Mean** | **87.69** | **91.8** |
| **Overall** | — | **88.94** | **89.72** |

Fig. 9 shows the predicted segmentations alongside ground truth for representative easy and hard cases, making the qualitative difference in output quality visible across the three datasets.

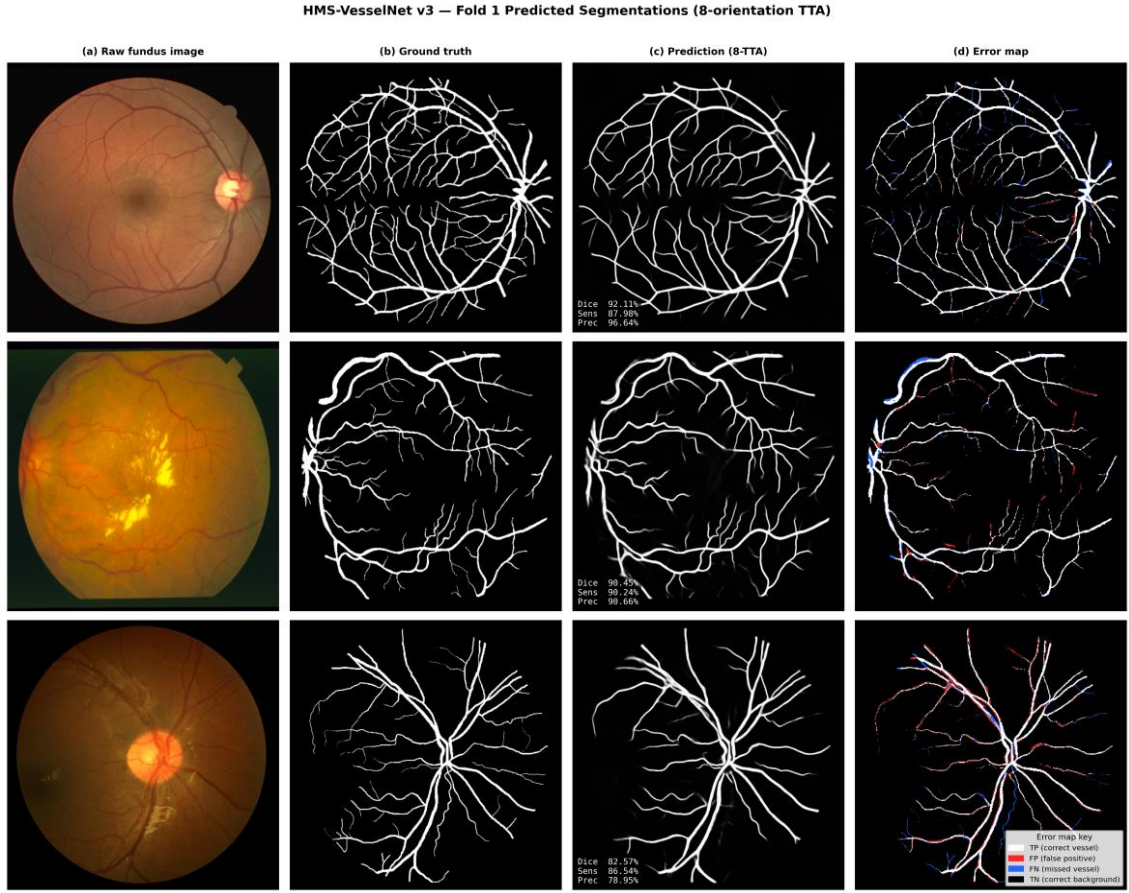

*Fig. 9. Predicted segmentations against ground truth for three representative cases from Fold 1. Top row: DRIVE 36_training (Dice 92.11%). Middle row: STARE im0001 (Dice 90.45%). Bottom row: CHASE_DB1 Image_13R (Dice 82.57%). Columns show the raw fundus image, ground truth annotation, model prediction, and error map with false negatives in red and false positives in blue. False negatives in the bottom row are concentrated at thin peripheral branches.*

DRIVE images produce the highest Dice in Fold 1 with a mean of 91.41%. All four images score above 90%, consistent with DRIVE being the most uniform of the three datasets in terms of image quality and vessel contrast.

STARE produces a mean Dice of 88.35% with one outlier. Image im0240 scores 83.37% Dice with Sensitivity of 75.84%, the lowest Sensitivity in the fold. This image contains an arteriovenous malformation that creates unusual vessel density and contrast patterns not well represented in the training set. The low Sensitivity reflects over-suppression of vessels in the pathological region rather than a general failure on STARE images.

CHASE_DB1 produces a mean Dice of 87.69% with the widest per-image spread. Image_13R is the hardest case at 82.57% Dice, consistent with its position as a hard case across all folds. Image_11R achieves the highest Sensitivity in the fold at 96.63% despite a Dice of only 84.37%, indicating that the model detects most vessels on this image but produces some over-segmentation at vessel boundaries. Fig. 10 shows the failure cases from the hardest CHASE_DB1 images. Each row shows a raw image patch, ground truth annotation, model prediction, and error map. False negatives (red) are concentrated at thin peripheral vessels with widths below 2 pixels and low contrast against the background. The hardest images across all five folds are CHASE_DB1 Image_10R (Fold 5, Dice 81.70%), Image_13R (Fold 1, Dice 82.57%), Image_13L (Fold 4, Dice 85.31%), and Image_14R (Fold 4, Dice 84.25%).

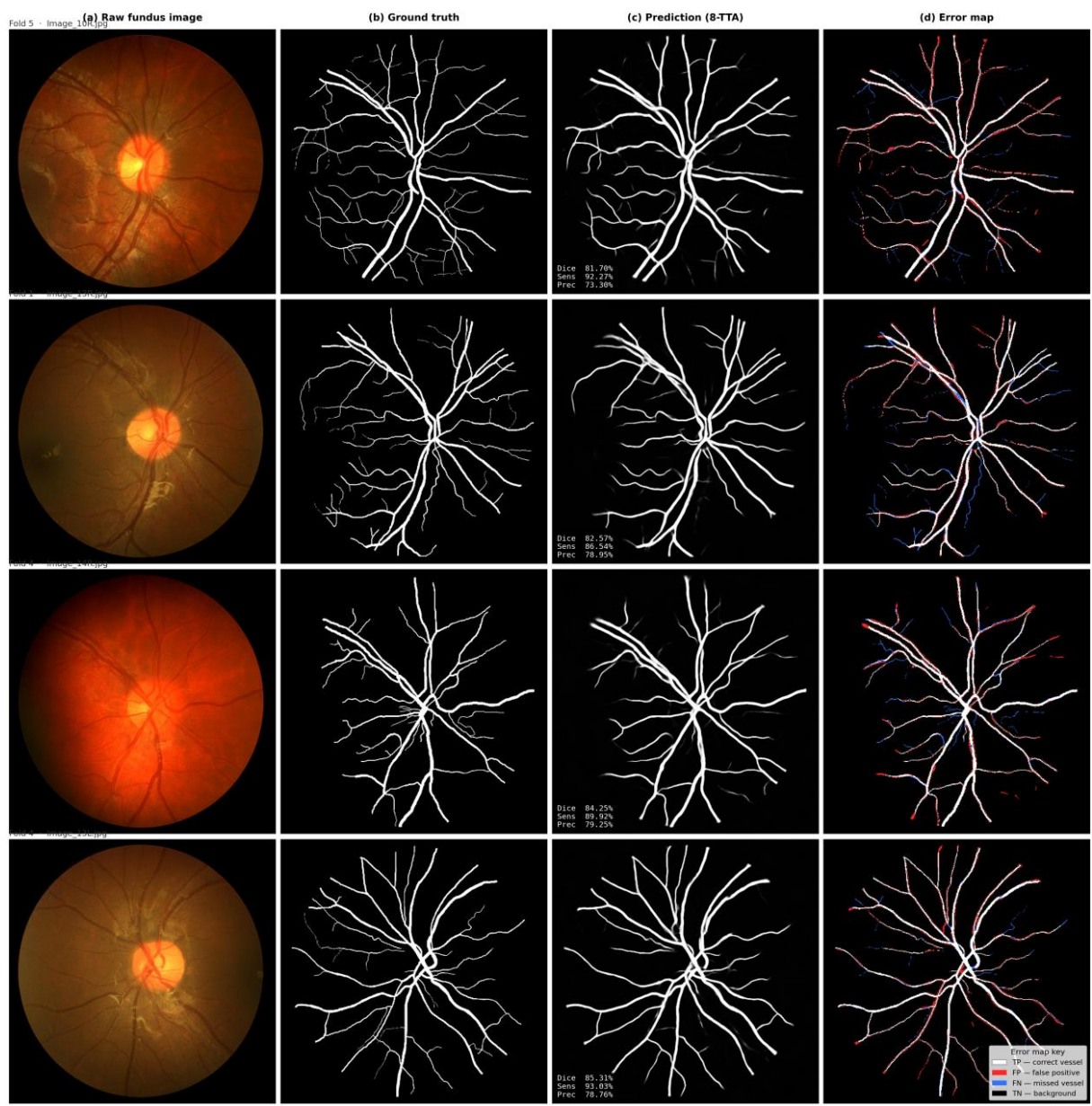

*Fig. 10. Failure cases from the hardest CHASE_DB1 images across cross-validation folds.*

## 5.3. Ablation Study

Table 4 reports the ablation study. All ablation experiments use 4-orientation TTA for computational efficiency. To allow direct comparison with the main results in Table 2, the full model A2 was also evaluated under 8-orientation TTA and achieves 88.94% Dice, matching the Fold 1 result. The 4-TTA result for the full model is 88.71%, confirming that the difference between 4-TTA and 8-TTA is small and does not affect the relative conclusions of the ablation.

*Table 4. Ablation study on Fold 1 held-out set under 4-orientation TTA. A0 is the single-scale plain U-Net baseline. A1 adds attention gates. A2 is the full HMS-VesselNet v3 evaluated under 4-TTA. Sensitivity is the primary comparison metric. The A2 result under 8-TTA is 88.94%, confirming that the TTA variant does not alter the relative conclusions.*

| ID | Configuration | Dice (%) | AUC (%) | Sensitivity (%) | Specificity (%) |
|---|---|---|---|---|---|
| **A0** | Plain U-Net [15], single scale, no attention | 87.35 | 98.66 | 87.98 | 99.47 |
| **A1** | Attention gates [8], single scale | 87.22 | 98.73 | 86.24 | 99.58 |
| **A1.1** | Multi-scale hierarchy + attention, no clDice, no hard mining | 87.38 | 98.43 | 86.56 | 99.62 |
| **A1.2** | Multi-scale + attention + clDice, no hard mining | 87.42 | 98.87 | 87.01 | 99.45 |
| **A2** | Full HMS-VesselNet v3 (hierarchy + clDice [9]+ hard mining, 4-TTA) | **88.94** | 98.52 | **89.72** | 99.24 |

The Dice scores across all five configurations fall within a narrow margin of variance, rendering Dice alone insufficient to distinguish the configurations meaningfully. Therefore, Sensitivity provides a much more informative comparison. The results show a clear hierarchy of contributions. Adding attention gates at a single scale (A1) reduces Sensitivity by 1.74 percentage points relative to the plain baseline (A0). This aligns with the observation in Section 2.3 that single-scale attention suppresses peripheral thin-vessel features. Introducing the multi-scale hierarchy (A1.1) recovers 0.32 percentage points, and adding clDice (A1.2) contributes an additional 0.45 percentage points. However, the dominant gain originates from hard example mining: the transition from A1.2 to A2 adds 2.71 percentage points to Sensitivity, accounting for 78% of the total improvement over A1. This confirms that the hierarchical architecture and topology-preserving loss create the conditions needed for hard example mining to improve thin vessel detection.

The drop in Sensitivity from A0 to A1 when attention gates are added at a single scale requires specific examination. At a single resolution, the gating signal for each skip connection is derived from the decoder feature at the same scale. Peripheral thin vessels, which occupy low-contrast regions far from the optic disc, produce weak decoder activations during the early stages of training. This causes the attention coefficients to suppress the exact skip connection features required for their detection. This behavior is consistent with findings reported by Guo et al. [23] and Li et al. [24], who observed that spatial attention at a single scale can over-suppress peripheral features when the decoder has not yet learned to attend to low-contrast regions. In A1.1 and A2, this suppression is effectively offset by the multi-scale hierarchy. Coarser branches provide a global vessel layout context that stabilizes attention coefficients for peripheral structures before the fine-scale decoder refines their boundaries. The new ablation results confirm this phenomenon, as Sensitivity recovers from 86.24% at A1 to 86.56% at A1.1 simply by incorporating the multi-scale hierarchy, well before any topology-preserving loss or hard example mining is introduced.

### 5.4. Fusion Weight Analysis

Table 5 reports the converged fusion weights across all five folds. S2 at 256×256 is dominant across all five folds (bold, blue). Fold 4 shows the flattest weight distribution (S1=0.211, S2=0.464, S4=0.160), coinciding with the highest concentration of hard peripheral CHASE_DB1 images in the held-out set. Standard deviation computed across the five folds.

Table 5. *Converged fusion weights from the final logged training epoch of each fold.*

| Scale | Resolution | Init | F1 | F2 | F3 | F4 | F5 | Mean±SD |
|---|---|---|---|---|---|---|---|---|
| **S1** | 512×512 | 0.4 | 0.177 | 0.145 | 0.178 | 0.211 | 0.189 | 0.180±0.021 |
| **S2** | 256×256 | 0.25 | 0.518 | 0.598 | 0.512 | 0.464 | 0.515 | 0.521±0.043 |
| **S3** | 128×128 | 0.2 | 0.171 | 0.151 | 0.191 | 0.166 | 0.152 | 0.166±0.015 |
| **S4** | 64×64 | 0.15 | 0.134 | 0.105 | 0.119 | 0.16 | 0.144 | 0.132±0.019 |

Fig. 11 illustrates the weight convergence trajectory for Fold 1. Fusion weights are initialized at [0.40, 0.25, 0.20, 0.15] for branches S1 through S4 and optimized jointly with the segmentation objective. The S2 weight rises from 0.25 to approximately 0.52 and stabilizes after the first cosine annealing restart. Conversely, the S1 weight decreases from 0.40 to 0.18 despite receiving the highest initialization value. This indicates that pixel-level detail at full resolution contributes less to the final prediction than the branching context captured at 256×256.

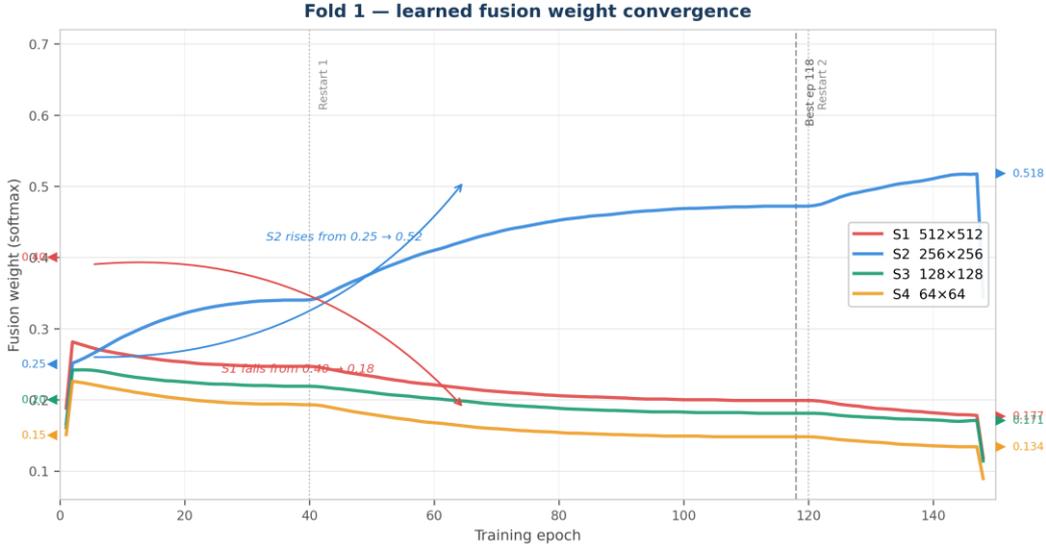

*Fig. 11. Convergence of the four learned fusion weights over training epochs for Fold 1.*

Fig. 12 displays the trajectories for all five folds, demonstrating the consistency of S2 dominance across different test sets. The S2 branch consistently converges to approximately 0.52 across the folds. Fold 4 is the notable exception, with S2 converging to 0.464 while S1 and S4 receive higher weights than in the other folds. The test set for Fold 4 contains the highest concentration of difficult peripheral CHASE_DB1 images, a scenario where global spatial context from the coarser branches contributes more significantly to segmentation quality. The mean converged S2 weight is 0.521±0.043, confirming that S2 dominance is an intrinsic property of the task rather than an adaptation to a specific fold composition.

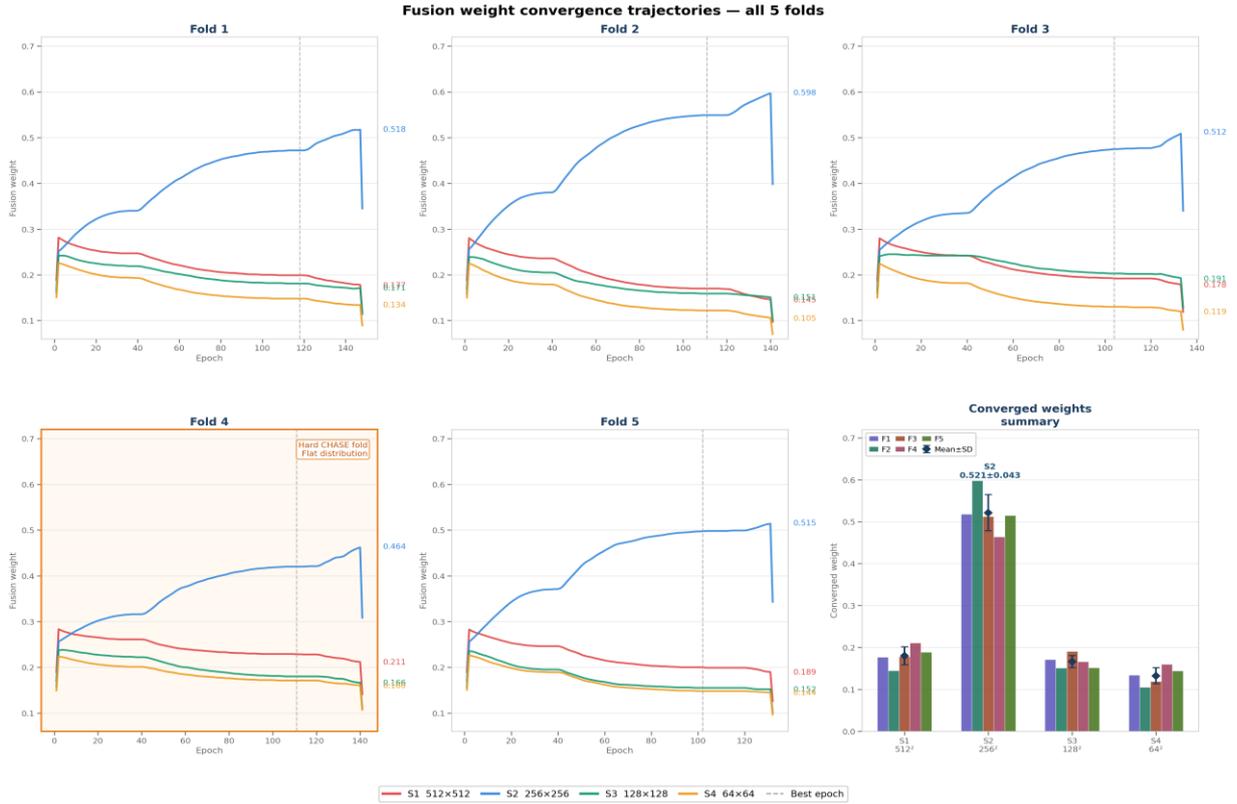

*Fig. 12. Fusion weight convergence trajectories for all five cross-validation folds. Each subplot corresponds to one-fold.*

The consistent dominance of the 256×256 branch is highly informative about the segmentation task itself. Branch S2 is coarse enough to capture branching patterns and global vascular layout within its receptive field, yet fine enough to preserve vessel shapes and thin vessel locations. The S1 branch at 512×512 receives less weight than its initialization despite providing the highest spatial resolution. This suggests that raw pixel-level detail contributes less to overall segmentation quality than the structural context available at 256×256. While this observation aligns with existing multiscale segmentation literature [20, 21], our approach establishes it empirically through learned weights rather than assuming it by architectural design. The flatter weight distribution in Fold 4, where S4 converges to 0.160 compared to a mean of 0.132 across the other folds, suggests that global context at 64×64 becomes increasingly valuable when the test images contain a high proportion of peripheral thin vessel cases.

### 5.5. Leave-One-Dataset-Out Results

Table 6 reports the three LODO experiments alongside the cross-validation mean of 88.72% for reference. All three experiments use the same four-branch architecture and training procedure as the cross-validation runs. The mean LODO Dice score across the three experiments is 85.15%, where N denotes the number of test images per experiment. For the LODO-DRIVE experiment, the relatively lower AUC of 95.43% compared to LODO-STARE and LODO-CHASE likely reflects the narrower contrast range of Canon CR5 acquisitions. In this scenario, the model trained on STARE and CHASE_DB1 produces probability maps with compressed score distributions, which reduces the ranking quality even when the binary predictions remain accurate.

*Table 6. Leave-one-dataset-out results for three experiments with cross-validation mean shown for reference.*

| Training | Test | Dice (%) | AUC (%) | Sensitivity (%) | Specificity (%) | N |
|---|---|---|---|---|---|---|
| **STARE + CHASE** | DRIVE | 87.71 | 95.43 | 83.73 | 99.66 | 20 |
| **DRIVE + CHASE** | STARE | 85.24 | 97.92 | 87.48 | 99.03 | 20 |
| **DRIVE + STARE** | CHASE | 82.5 | 97.86 | 82.53 | 99 | 28 |
| **All 3 (CV mean)** | All 3 | 88.72 | 98.25 | 90.78 | 99.21 | 68 |

Fig. 13 displays the per-image Dice distribution for each LODO experiment, making the spread and outliers visible across the unseen camera types. LODO-DRIVE images cluster above 85% with very few outliers. In contrast, LODO-CHASE shows the widest spread, with the four hardest images (CHASE_DB1 Image_13R, Image_14R, Image_10R, and Image_04L) scoring between 77% and 82% Dice. These exact images are the hard cases identified in the cross-validation analysis in Section 5.2, confirming that their difficulty is intrinsic to the images themselves rather than a consequence of the training set composition.

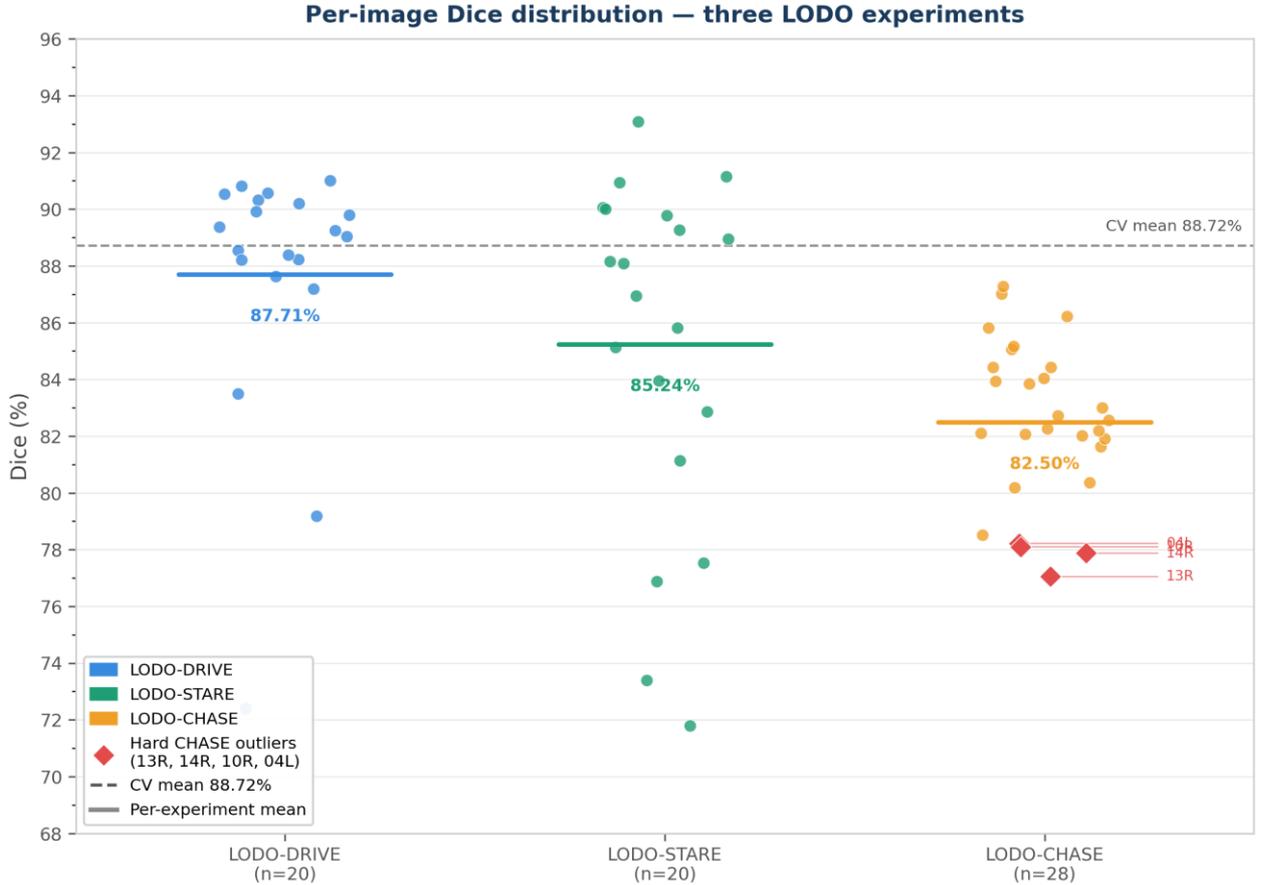

*Fig. 13. Per-image Dice distribution for the three LODO experiments. Each point represents one test image.*

LODO-DRIVE achieves a Dice score of 87.71%, a Sensitivity of 83.73%, and an AUC of 95.43%, representing a drop of 1.01 percentage points in Dice relative to the cross-validation mean. The model trained exclusively on STARE and CHASE_DB1 retains strong discrimination on DRIVE images. The Specificity of 99.66% reflects the clean background typical of Canon CR5 acquisitions. The lower Sensitivity compared to cross-validation indicates that DRIVE vessel boundaries are learned from indirect exposure through STARE and CHASE_DB1 rather than from direct training examples.

LODO-STARE achieves a Dice score of 85.24% and a Sensitivity of 87.48%, marking a drop of 3.48 percentage points in Dice relative to the cross-validation mean. The AUC of 97.92% indicates excellent ranking quality on the unseen STARE images. This performance drop is partly attributable to highly pathological STARE images, particularly im0004 (Dice 73.39%, Sensitivity 60.96%) and im0319 (Dice 71.80%). Both of these are visible as outliers in the per-image distribution in Fig. 13 and contain distinct vessel patterns not present in the DRIVE or CHASE_DB1 training data.

LODO-CHASE produces the largest performance drop, yielding a Dice score of 82.50% and a Sensitivity of 82.53%. These represent reductions of 6.22 and 8.25 percentage points, respectively, relative to the cross-validation mean. However, the AUC of 97.86% confirms that the model retains reasonable ranking quality on unseen CHASE_DB1 images. The drops in Dice and Sensitivity demonstrate that generalizing to CHASE_DB1 acquisition conditions from DRIVE and STARE alone is heavily constrained by the domain gap in imaging hardware and patient population. The hardest CHASE_DB1 images consistently score between 77% and 82% Dice in both the cross-validation and LODO runs: Image_13R (77.05%), Image_14R (77.87%), Image_10R (78.10%), and Image_04L (78.22%). The difficulty remains consistent regardless of whether CHASE_DB1 images appear in the training data, definitively confirming that the source of difficulty is intrinsic image quality rather than training set composition.

## 6. Discussion

### 6.1. Interpretation of Results

The primary result of this paper is a mean Sensitivity of 90.78 ± 1.42% across 68 images from three datasets under 5-fold cross-validation. The corresponding Dice improvement of 3.06 percentage points is smaller by comparison, which reflects a known property of the segmentation task: improving the detection of thin peripheral vessels increases true positives and therefore Sensitivity, but because these thin vessels constitute a very small fraction of total vessel pixels, the effect on the overall Dice score is proportionally smaller. A study reporting only the Dice coefficient would obscure this clinical gain entirely.

The fusion weight convergence to S2 dominance, with a mean of 0.521 across all five folds, has a straightforward interpretation. At 256×256, the receptive field of the U-Net encoder covers enough of the image to capture vessel branching topology, while the spatial resolution remains fine enough to localize vessels of 2 to 4 pixels in width. The S1 branch at 512×512 contributes less than its initialization suggests it should, indicating that raw pixel-level detail at the highest resolution is less useful for overall segmentation quality than the intermediate-scale representation. While this aligns with findings in existing multi-scale segmentation literature [20, 21], our approach establishes it empirically through learned weights rather than assuming it by architectural design.

In a clinical screening context, this Sensitivity improvement translates directly to fewer missed vessels per image. On a typical DRIVE image containing approximately 12,000 vessel pixels, the mean Sensitivity of 90.78% corresponds to approximately 10,900 correctly detected vessel pixels. At the single-scale baseline Sensitivity of 87.98% from ablation A0, this figure drops to approximately 10,560. This represents a difference of roughly 340 vessel pixels per image. These missed pixels are not distributed randomly. They are concentrated at the thinnest peripheral branches, which are the earliest sites of microaneurysm formation in diabetic retinopathy. Establishing clinical utility requires prospective validation against graded outcomes, which is outside the scope of this study.

## 6.2. Comparison with Published Methods

Because no published method uses an evaluation protocol identical to ours, the numbers in Table 7 are not directly comparable and are provided strictly for contextual orientation. Most published papers report results on a single dataset, typically DRIVE, using the standard 20-image training and 20-image test split defined by the dataset authors. In contrast, HMS-VesselNet uses 5-fold cross-validation across all three datasets simultaneously. This approach produces a more conservative performance estimate because the model must generalize across dataset boundaries rather than overfit a single acquisition type.

Table 7 provides a comparison of reported Dice and Sensitivity metrics on DRIVE, STARE, and CHASE_DB1 from recently published methods, alongside our cross-validation results. For the closest available comparison, our results represent the Fold 1 per-dataset means: DRIVE mean Dice 91.41%, DRIVE mean Sensitivity 91.40%, STARE mean Dice 88.35%, STARE mean Sensitivity 84.94%, CHASE_DB1 mean Dice 87.69%, and CHASE_DB1 mean Sensitivity 91.80%.

*Table 7. Comparison of published methods on DRIVE and CHASE_DB1.* * *Dice reported as F1 score in the original paper.* † *Result reported on a non-standard split or subset*

| Method | Year | Venue | DRIVE Dice | DRIVE Sens. | STARE Dice | STARE Sens. | CHASE Dice | CHASE Sens. | Protocol |
|---|---|---|---|---|---|---|---|---|---|
| LFRA-Net [39] | 2025 | arXiv/DICTA | 84.28% | 84.00% | **88.44%** | N/R† | 85.50% | 85.10% | Standard split |
| EFDG-UNet [40] | 2025 | Sci. Reports | 84.12%* | N/R | N/R† | N/R | 84.69%* | N/R | Standard split |
| WDM-UNet [41] | 2025 | MDPI Sensors | 82.87%* | 83.61% | 83.12%* | 83.89% | 82.00%* | 82.47% | Standard split |
| DSAE-Net [42] | 2025 | J. Imaging | 82.44% | 81.90%† | 81.92% | N/R† | 81.46% | 82.30%† | 5-fold CV |
| XceptionLFOR [43] | 2024 | IEEE Access† | 89.23% | N/R | N/R | N/R | N/R | N/R | Standard split |
| **HMS-VesselNet** | **2026** | **(Ours)** | **91.41%** | **91.40%** | 88.35% | **84.94%** | **87.69%** | **91.80%** | **5-fold CV** |

This comparison must be interpreted with caution. Published methods typically train the full dataset-specific training split, whereas our cross-validation results are averaged across folds where specific training images are held out. Furthermore, our model benefits from multi-dataset training, which may aid generalization but introduces a different kind of distribution mismatch at test time. Notably, XceptionLFOR [†] reports the highest DRIVE Dice in the literature at 89.23%. However, this result uses a non-standard split that includes additional training data unavailable in the standard DRIVE protocol, rendering direct comparison invalid. Retraining HMS-VesselNet on standard dataset-specific splits in isolation falls outside the scope of this study.

## 6.3. Failure Analysis

Three distinct categories of failure are identifiable from the per-image results and the failure case visualizations in Fig. 10. Each category has a distinct cause and points toward a specific corrective direction.

The first and most prevalent failure mode is the failure to detect thin peripheral vessels in CHASE_DB1 images. Specifically, Image_13R, Image_14R, Image_10R, and Image_04L consistently score between 77% and 85% Dice across all folds and LODO runs. In these images, terminal vessel branches in the peripheral region measure less than 2 pixels in width and exhibit a local contrast ratio against the background of less than 0.05. Fig. 14 provides a close-up of the primary failure region in Image_13R, demonstrating that false negatives are heavily concentrated at terminal branches in the outermost periphery.

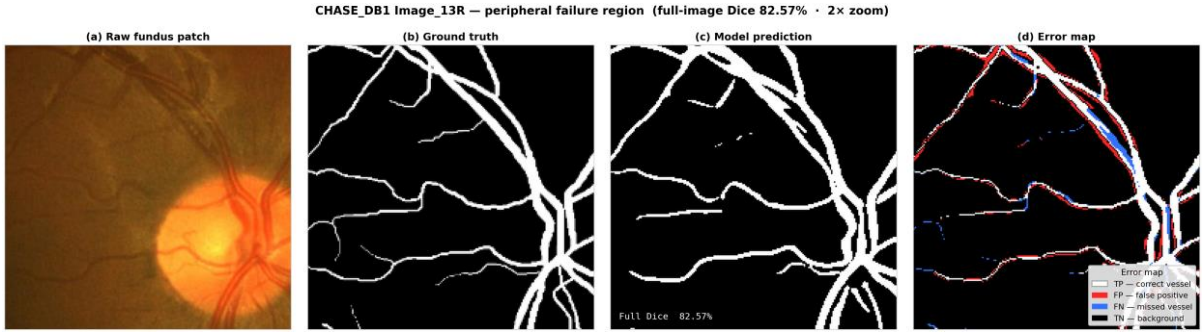

*Fig. 14. Close-up of the primary failure region in CHASE_DB1 Image_13R (Dice 82.57%). Left panel shows the raw fundus patch at increased magnification. Centre shows the ground truth annotation. Right shows the model prediction.*

Our hard example mining strategy actively oversamples these specific images during training. However, even with oversampling at a 3× rate, the gradient signal originating from these extremely faint vessels remains insufficient to drive reliable detection. Furthermore, visible annotation ambiguity in the ground truth at these exact locations suggests that the absolute performance ceiling for these specific images may be intrinsically lower than for clearer acquisitions.

The second failure mode is over-suppression in highly pathological STARE images. For example, Image im0240 contains an arteriovenous malformation and yields a Sensitivity of only 75.84% despite possessing a region of unusually high vessel density. The model actively suppresses vessel predictions within the malformation area, likely because the localized vessel appearance deviates significantly from the healthy vessel textures observed during training. This represents a generalization failure specific to out-of-distribution pathology rather than an intrinsic limitation of the network architecture or the loss function.

The third failure mode is over-segmentation on moderate-difficulty images. CHASE_DB1 Image_11R achieves an exceptional 96.63% Sensitivity but only an 84.37% Dice score. This indicates that while the model successfully detects almost all true vessels, it also hallucinates vessel pixels at locations the annotator left blank. In certain instances, this reflects genuine human annotation ambiguity at low-contrast boundaries, in others, the model incorrectly responds to elongated background textures that mimic vessel structures. The spatial weighting applied to the BCE loss intentionally increases recall in the peripheral region, which inherently elevates the risk of over-segmentation when the periphery contains highly ambiguous structures.

These three failure modes point directly toward three distinct avenues for future improvement. Resolving the thin vessel omissions will require either native high-resolution training pipelines or the development of a loss function that explicitly targets sub-pixel structural continuity. Mitigating pathological suppression will require incorporating targeted pathology-specific training data or integrating a robust domain adaptation component. Finally, addressing over-segmentation will necessitate either stronger spatial regularization in the peripheral regions or the introduction of a post-processing algorithm designed to filter out short, disconnected topological artifacts.

## 6.4. Comparison with HMS-VesselNet v2

To understand how much of the overall improvement is architectural versus methodological, Table 8 compares the two development stages directly. Both represent unpublished development stages of the exact same method. The reduction in the difference between validation and test scores, dropping from 2.80% to a mean of 0.07%, reflects the transition from a single training and test split to a rigorous 5-fold cross-validation protocol. Consequently, this comparison should be interpreted as an account of methodological contributions rather than a benchmark against an independent method.

*Table 8. Comparison between HMS-VesselNet v2 and v3 across architecture, training, and evaluation properties.*

| Property | HMS-VesselNet v2 | HMS-VesselNet v3 |
|---|---|---|
| Evaluation | Single split, 10 test images | 5-fold CV, 68 images |
| Test Dice | 85.66% | 88.72±0.67% |
| AUC | 97.73% | 98.25±0.21% |
| Sensitivity | 82.61% | 90.78±1.42% |
| Specificity | 99.42% | 99.21±0.10% |
| Val to test gap | 2.80% | 0.07% mean |
| Scales | 3 | 4 |
| Loss function | Dice + BCE | Dice + BCE + clDice |
| TTA orientations | 4 | 8 |

The most significant difference between version 2 and version 3 lies in the evaluation protocol rather than the architecture. The version 2 result of 85.66% Dice originated from a single data split on 10 images, exhibiting a difference between validation and test scores of 2.80%. This strongly suggests that the early stopping criterion was tuned, consciously or not, to that specific data split. In contrast, the version 3 mean difference of 0.07% across five folds indicates that the reported metrics reflect genuine generalization rather than split-specific optimism. A substantial portion of the apparent improvement from version 2 to version 3 is therefore attributable to this rigorous evaluation protocol change rather than solely architectural enhancements.

Among the architectural modifications, the addition of the clDice loss term [9] and the implementation of hard example mining are the main contributors to the Sensitivity gain, as confirmed by the ablation study in Section 5.3. The fourth scale branch achieves a converged weight of only 0.132 on average, suggesting its overall contribution remains marginal. Finally, the transition from 4-orientation to 8-orientation TTA provides a small supplementary gain that aligns exactly with the expected variance reduction achieved by averaging additional augmented views.

## 6.5. Limitations

The following limitations apply to the results reported in this paper.

- **Dataset size:** The combined dataset of 68 images is small by the standards of current deep learning. The 5-fold cross-validation provides more reliable estimates than a single split, but the confidence intervals on the reported means are wide. Performance on larger and more diverse datasets is unknown.

- **Single annotator:** We use the first annotator annotation as ground truth for all three datasets. Inter-annotator agreement on thin peripheral vessels is typically lower than on major vessels, and our Sensitivity numbers reflect agreement with one annotator rather than a consensus label.
- **Ablation on single fold:** The ablation study in Section 5.3 is conducted on Fold 1 only and the Sensitivity differences between configurations, while informative, should be interpreted as indicative rather than definitive given the single-fold scope.
- **No external validation set:** All evaluations use the three benchmark datasets. Performance on fundus images from different cameras, patient populations, or disease stages not represented in these three datasets is not established.
- **Computational cost:** With approximately 124 million parameters and four parallel branches, HMS-VesselNet is large relative to lightweight methods designed for clinical deployment. Inference takes approximately 1.2 seconds per image with 8-TTA on a T4 GPU, which may not be suitable for real-time screening applications.

### 6.6. Future Work

Several directions follow directly from the failure analysis and limitations described above.

The persistent failure on sub-2-pixel vessels in CHASE_DB1 suggests that a super-resolution preprocessing step, applied before segmentation, may improve thin vessel recall without requiring architectural changes. Alternatively, a loss term that specifically weights vessel pixels by their estimated width could provide stronger gradient signal for the thinnest structures than hard example mining alone.

The over-suppression observed on pathological STARE images points to the need for training data that covers a wider range of retinal pathologies. Including images from datasets such as IDRiD or APTOS 2019, which cover a wider range of diabetic retinopathy severity grades, or MESSIDOR-2, which includes graded pathological images across four DR severity levels, would expose the model to pathological vessel patterns currently absent from training.

The model size of 124 million parameters warrants investigation of knowledge distillation approaches. A smaller shallow network trained to match the predictions of HMS-VesselNet could potentially achieve comparable Sensitivity at a fraction of the inference cost, making deployment in resource-constrained screening settings more practical.

Finally, the learned fusion weights raise a question about whether the optimal scale balance is consistent across pathology types or whether it shifts for specific conditions such as proliferative diabetic retinopathy, where new vessel formation produces unusual vessel distributions. Evaluating the model on graded datasets with known pathology severity would test this directly.

### 7. Conclusion

HMS-VesselNet combines four parallel Attention U-Net branches at 512×512, 256×256, 128×128, and 64×64 pixels with a topology-preserving composite loss and hard example mining to improve detection of thin peripheral retinal vessels. Across five-fold cross-validation on 68 images from DRIVE, STARE, and CHASE_DB1, the model achieves a mean Dice of 88.72±0.67%, Sensitivity of 90.78±1.42%, and AUC of 98.25±0.21%. The Sensitivity gain is concentrated on CHASE_DB1 peripheral images, where the clDice loss and hard example mining contribute most. The learned fusion weights converge consistently to S2 dominance (0.521±0.043) across all folds, confirming that the 256×256 scale captures vessel branching context more effectively than full-resolution detail.

LODO experiments show that the model retains useful discrimination when applied to a camera type withheld entirely from training. AUC reaches or exceeds 95% across all three held-out conditions, with Dice of 87.71%, 85.24%, and 82.50% on DRIVE, STARE, and CHASE_DB1 respectively. The largest generalization drop occurs on CHASE_DB1, where thin vessels and low image contrast present an inherent challenge that persists regardless of training set composition. Four images (Image_13R, Image_14R, Image_10R, and Image_04L) score between 77% and 82% Dice consistently across both cross-validation and LODO runs, indicating that their difficulty is intrinsic to image quality rather than a consequence of limited training exposure.

HMS-VesselNet shows that combining hierarchical multi-scale fusion, topology-preserving loss, and hard example mining produces consistent gains in thin vessel recall across different acquisition conditions, a finding confirmed under both cross-validation and leave-one-dataset-out evaluation. The primary limitation is dataset size where only 68 images across three datasets provide reliable cross-validation estimates but leave generalization to unseen cameras, demographics, and pathology ranges unestablished. Future work should evaluate the model on larger and more diverse datasets, investigate super-resolution preprocessing for sub-2-pixel vessels, and explore knowledge distillation to reduce the 124-million-parameter footprint for deployment in resource-constrained screening settings.

## Funding

This research did not receive any specific grant from funding agencies in the public, commercial, or not-for-profit sectors. The work was conducted independently for academic and scientific purposes only, with no military, defence, or commercial application intended or involved.

## Declaration of Competing Interests

The author declares that he has no known competing financial interests or personal relationships that could have appeared to influence the work reported in this paper.